\documentclass[prl,aps,superscriptaddress,preprint,floatfix]{revtex4-1}

\usepackage{graphicx}
\usepackage{verbatim}
\usepackage{mathrsfs}
\usepackage{amsmath,amsfonts,amssymb}
\usepackage{wrapfig}
\usepackage{graphicx}
\usepackage{bbm}
\usepackage{graphics}

\def\3{2.8in}
\def\2{2.5in}
\def\4{3.0in}

\def \beq {\begin{equation}}
\def \eeq {\end{equation}}
\begin{document}

\title{Observation of metallic surface states in the strongly correlated\\*Kitaev-Heisenberg candidate Na$_{2}$IrO$_{3}$}

\author{Nasser~Alidoust}\affiliation {Joseph Henry Laboratory, Department of Physics, Princeton University, Princeton, New Jersey 08544, USA}

\author{Chang~Liu}\affiliation {Joseph Henry Laboratory, Department of Physics, Princeton University, Princeton, New Jersey 08544, USA}

\author{Su-Yang~Xu}\affiliation {Joseph Henry Laboratory, Department of Physics, Princeton University, Princeton, New Jersey 08544, USA}

\author{Ilya~Belopolski}\affiliation {Joseph Henry Laboratory, Department of Physics, Princeton University, Princeton, New Jersey 08544, USA}

\author{Tongfei~Qi} \affiliation{Center for Advanced Materials, Department of Physics and Astronomy, University of Kentucky, Lexington, Kentucky 40506, USA}

\author{Minggang~Zeng}\affiliation {Graphene Research Centre and Department of Physics, National University of Singapore, Singapore 117542}

\author{Madhab~Neupane}\affiliation {Joseph Henry Laboratory, Department of Physics, Princeton University, Princeton, New Jersey 08544, USA}

\author{Guang~Bian}\affiliation {Joseph Henry Laboratory, Department of Physics, Princeton University, Princeton, New Jersey 08544, USA}

\author{Yu-Tzu~Liu}\affiliation {Graphene Research Centre and Department of Physics, National University of Singapore, Singapore 117542}

\author{Stephen~D.~Wilson}\affiliation {Materials Department, University of California, Santa Barbara, California 93106, USA}

\author{Hsin~Lin}\affiliation {Graphene Research Centre and Department of Physics, National University of Singapore, Singapore 117542}

\author{Arun~Bansil}\affiliation {Department of Physics, Northeastern University, Boston, Massachusetts 02115, USA}

\author{Gang~Cao} \affiliation{Center for Advanced Materials, Department of Physics and Astronomy, University of Kentucky, Lexington, Kentucky 40506, USA}

\author{M.~Zahid~Hasan}\affiliation {Joseph Henry Laboratory, Department of Physics, Princeton University, Princeton, New Jersey 08544, USA}


\begin{abstract} We report high-resolution angle-resolved photoemission spectroscopy measurements on the honeycomb iridate Na$_{2}$IrO$_{3}$. Our measurements reveal the existence of  a metallic surface band feature crossing the Fermi level with nearly linear dispersion and an estimated surface carrier density of 3.2 $\times$ 10$^{13}$ cm$^{-2}$, which has not been theoretically predicted or experimentally observed, and provides the first evidence for metallic behavior on the boundary of this material, whereas the bulk bands exhibit a robust insulating gap. We further show the lack of theoretically predicted Dirac cones at the $\overline{M}$ points of the surface Brillouin zone, which confirms the absence of a stacked quantum spin Hall phase in this material. Our data indicates that the surface ground state of this material is exotic and metallic, unlike as predicted in theory, and establishes Na$_{2}$IrO$_{3}$ as a rare example of a strongly correlated spin-orbit insulator with surface metallicity.
\end{abstract}

\date{\today}
\maketitle

Over the last few years, iridium oxides (iridates) have received considerable attention in condensed matter physics \citep{Cao2002, Moon2008, Kim2008, Jackeli2009, Shitade2009, Machida2010, Chaloupka2010, Pesin2010, Wan2011, Kimchi2014, Kim2012}. This has been mainly due to their rich ground state phase diagrams, which involve spin-orbit coupling, on-site Coulomb interactions, and direct and indirect hopping between different elements. All of these various types of interactions have comparable energy scales, creating a unique platform to study their interplay in real materials. For instance, in many of the iridates, electron correlations lead to a Mott or many-body insulating ground state. Furthermore, spin-orbit coupling, an important interaction in iridates, has come to play an increasingly important role in condensed matter physics since the discovery of topological insulators \citep{Hasan2010, Qi2011, Hsieh2008a, Hsieh2008b, Xia2009, Chen2010, Hsieh2009, Xu2012}, because it leads to the non-trivial topology of the electronic band structure in this class of materials. However, almost all topological insulators discovered to date are weakly correlated materials, and only recently some experimental evidence for the existence of topological phases in the strongly correlated Kondo insulator SmB$_{6}$ has been found \citep{Dzero2010, Neupane2013, Jiang2013}. The \textit{A}$_{2}$IrO$_{3}$ honeycomb iridate is one of the recently debated series among the iridate compounds, and in this series Na$_{2}$IrO$_{3}$ is especially interesting. There have been proposals of a possible Kitaev-Heisenberg magnetic ground state in this compound and the existence of a quantum spin liquid phase \citep{Chaloupka2010, Liu2011, Choi2012, Ye2012, Chaloupka2013}. Moreover, and more relevant to our present work, this material has been proposed to possess topologically non-trivial band structure with protected metallic surface states \citep{Shitade2009, Pesin2010, Sohn2013, Kim2012}. Specifically, a quantum spin Hall (QSH) state has been predicted to exist in this compound, whose hallmark signature is the existence of Dirac cones at the $\overline{M}$ points of the Brillouin zone (edge Kramers' points, unlike Bi$_{2}$Se$_{3}$) in a stacked 3D configuration of this material \citep{Shitade2009, Kim2012}. It has experimentally been suggested that Na$_{2}$IrO$_{3}$ is a relativistic Mott insulator \citep{Singh2010, Comin2012, Gretarsson2013}. There have also been indications that this compound is a Slater insulator \citep{Kim2013}. However, the issue is not fully resolved yet and moreover, there has not been any experimental evidence supporting the emergence of any surface states in this intriguing Mott-like insulating phase. 

In this paper we report systematic ARPES studies on Na$_{2}$IrO$_{3}$ single crystals mapping the first Brillouin zone (BZ) entirely. We report the observation of a metallic linear-like surface band feature crossing the Fermi level at the $\overline{\Gamma}$ point (center of the surface BZ), providing the first evidence for metallic behavior on the surface boundary of this material, despite the fact that our experimentally observed bulk bands exhibit a robust gap. This is in contrast to the theoretical prediction of Dirac cones at the $\overline{M}$ points of the BZ. Therefore, our results suggest that the surface ground state electronic structure of this system is not in a stacked QSH phase as predicted theoretically, but rather reflects an exotic many-body metallic ground state on the surface. It is worth mentioning that measuring the electronic band structure of Na$_{2}$IrO$_{3}$ single crystals using ARPES is very challenging due to the small size of the samples and the low success rate in cleaving flat surfaces from them. However, we have managed to overcome these obstacles and obtain reproducible data. Our data indicates that the surface ground state of this material is metallic and more exotic than that predicted in theory, and establishes Na$_{2}$IrO$_{3}$ as a rare example of a strongly correlated spin-orbit insulator with surface metallicity. 

A self-flux method was used to grow single crystals of Na$_{2}$IrO$_{3}$ from off-stoichiometric quantities of IrO$_{2}$ and Na$_{2}$CO$_{3}$. Technical details can be found on the preparation of related iridate compounds in refs. \citep{Ye2012, Chikara2009, Laguna-Marco2010, Ge2011}. The studied crystals have typical sizes of ~2 mm $\times$ 1 mm $\times$ 0.5 mm. ARPES measurements were performed with incident photon energies of 60 - 100 eV,  and LEED measurements were performed with electron energies of 90 - 300 eV, both at beamline 4.0.3 of the Advanced Light Source (ALS) in the Lawrence Berkeley National Laboratory (LBNL). Samples were cleaved \textit{in situ} at 300 K in chamber pressure better than $5\times10^{-11}$ torr resulting in shiny surfaces. Energy resolution was better than 15 meV and momentum resolution was better than 1\% of the surface BZ. The measurements were also conducted at 300 K, which excludes the possibility of an unconventional low-temperature surface reconstruction reported on this material using scanning tunneling microscopy \citep{Lupke2014}. We note that we also found the ARPES signal to disappear entirely at those low temperatures where the STM measurements were performed. But at room temperature the metallic linear-like surface states that cross the Fermi level are clearly seen in our data (see Figs. in this paper).

The recently revised monoclinic structure of Na$_{2}$IrO$_{3}$ (space group: \textit{C2/m}, No. 12) consists of edge-sharing and \textbf{c}-stacking IrO$_{6}$ octahedra \citep{Choi2012, Ye2012}. Ir atoms form a nearly ideal honeycomb lattice with the centers occupied by Na. Different from the rhombohedral (ABC) stacking, an in-plane offset of successive Ir honeycombs gives rise to the monoclinic distortion; and one Na layer is sandwiched by the neighboring IrO$_{6}$ octahedra layers. The existence of trigonal distortions of oxygen ions in IrO$_{6}$ octahedra can split the t$_{2g}$ manifold. Na$_{2}$IrO$_{3}$ undergoes a magnetic phase transition into zigzag-antiferromagnetic (AFM) order at \emph{T}$_{\text{N}}$ $\sim$ 15 K. Therefore, crystal cells with eight (Fig. 1(a) (side view) and 1(b) (top view)) and two (Fig. 1(d) (side view) and 1(e) (top view)) formula units are used to describe the zigzag-AFM and non-magnetic phases, respectively. In Figs. 1(a) and 1(b) the long-range magnetic order is indicated by the local magnetic moments of Ir atoms, which adopt the zigzag spin configuration with magnetic moments along the \textbf{a} axis and AFM coupling between neighboring Ir zigzag-chains. The corresponding Brillouin zones for the magnetic and non-magnetic phases are illustrated in Figs. 1(c) and 1(f), respectively. A doubled surface unit cell, and thus a halved surface BZ, is adopted to consider the zigzag-AFM spin configuration. For comparison, as shown in Fig. 1(g), we plot the hexagonal surface BZ of the non-magnetic Na$_{2}$IrO$_{3}$ in green, along with that of the zigzag-AFM Na$_{2}$IrO$_{3}$ in red.

We conduct low-energy electron diffraction (LEED) measurements at various electron energies on the single crystals of Na$_{2}$IrO$_{3}$ used in our ARPES experiments. The results are shown in Fig. 1(h), and their analysis in Fig. S1. As revealed from our LEED measurements no detectable surface reconstruction is present in these crystals. We also measure the core level spectra of these crystals (Fig. 2(a)) and confirm the chemical composition of this compound by observing sharp Na and Ir peaks in the core level intensity plot, which is suggestive of the high quality of the Na$_{2}$IrO$_{3}$ samples used in our measurements.

In Fig. 2(b) we present a \emph{k}-\emph{E} ARPES intensity profile of Na$_{2}$IrO$_{3}$ over a wide binding energy scale measured with a synchrotron-based ARPES system using a photon energy of $h\nu$ = 90 eV at a temperature of \emph{T} $\sim$ 300 K. We observe two bulk bands in the binding energy range from the Fermi level to binding energy \emph{E}$_{\text{B}}$ = 6 eV: the first extending from a binding energy of $\sim$  0.5 eV to $\sim$ 3 eV, and the second from a binding energy of $\sim$ 3 eV to $\sim$ 6 eV. Our measurements indicate that the top of the bulk valence band is located approximately 500 meV below the Fermi level. The red dashed lines are guides-to-the-eye highlighting the M shape of the bulk valence band, indicative of its inverted nature. The stack plot of the corresponding energy distribution curves (EDCs) is shown in Fig. 2(c), which is in qualitative agreement with the previously reported EDCs of this material (see Fig. S2). We highlight that this is the first time that bands with clear dispersion have been observed on Na$_{2}$IrO$_{3}$ \citep{Comin2012}. These bands are relatively broader as compared to the ARPES data on many weakly interacting systems (such as graphene or Bi$_{2}$Se$_{3}$). We note that this is likely due to intrinsic correlation effects rather than lack of crystalline order of our samples, whose core level spectra demonstrate their high quality. The kinetic energy value of the Fermi level is determined by measuring that of the polycrystalline gold separately at the same photon energies and under the same experimental conditions (see Fig. 3(f)). 

Figure 2(d) and 2(e) show the measured momentum intensity maps at selected constant binding energies from the Fermi level \emph{E}$_{\text{F}}$ to \emph{E}$_{\text{B}}$ = 3 eV, with the high symmetry directions of the BZ, $\overline{\Gamma}-\overline{K}$ and $\overline{\Gamma}-\overline{M}$, marked on the last panel of Fig. 2(d).  At \emph{E}$_{\text{F}}$ we observe a circular pocket at the center of the BZ ($\overline{\Gamma}$ point), and upon going to larger binding energies (e.g. \emph{E}$_{\text{B}}$ = 2.5 eV) the constant binding energy momentum cuts evolve to six circular pockets at the corners of the BZ ($\overline{K}$ point). As demonstrated by the dashed guide-to-the-eye lines drawn at the place of each Fermi surface pocket in Fig. 2(e) these constant binding energy maps clearly demonstrate a hexagonal Brillouin zone, consistent with the six-fold symmetry of the (001) surface of this compound. 

An important feature of the electronic band structure of this compound revealed by our measurements is the observation of spectral intensities at the time-reversal invariant Kramers' $\overline{\Gamma}$ point of the BZ extending from the Fermi level to the top of the valence band (\emph{E}$_{\text{B}}$ $\sim$ 0.5 eV) as can be seen in Fig. 2(b). The purple rectangular dashed box in this figure highlights the momentum region in which these intensities are observed. These states form a metallic Fermi surface pocket at \emph{E}$_{\text{F}}$ as shown in the first panel of Fig. 2(d). In an attempt to understand these states in more detail, we next focus on studying the low-energy states of this material lying close to \emph{E}$_{\text{F}}$. 

In Fig. 3(a) and 3(b) we present \emph{k}-\emph{E} ARPES intensity profiles of the low-energy electronic states observed close to the Fermi level in this material, along the two high symmetry directions $\overline{\Gamma}-\overline{K}$ and $\overline{\Gamma}-\overline{M}$, respectively. These high symmetry directions are shown on the fifth panel of Fig 2(d). We focus on a narrow binding energy window between the Fermi level and \emph{E}$_{\text{B}}$ = 2 eV. These \emph{k}-\emph{E} intensity plots clearly show the existence of in-gap electronic states between the top of the valence band at E$_{\text{B}} \sim$ 0.5 eV and \emph{E}$_{\text{F}}$. For clarity, the momentum distribution curves (MDCs) at \emph{E}$_{\text{F}}$ are plotted at the top of each panel in purple, which further confirm the existence of finite intensities at \emph{E}$_{\text{F}}$ around the time-reversal invariant Kramers' $\overline{\Gamma}$ point. To study the dispersion of these metallic states in more detail, we perform a quantitative analysis by heuristically fitting the MDCs of these metallic in-gap states in the binding energy range between 50 meV and 450 meV around the $\overline{\Gamma}$ point using two Lorentzians (Fig. 3(c)). We then track the momentum positions of these peaks at the corresponding binding energies (purple dots in Fig. 3(d)), which reveals the nearly linear dispersion of these states (for details of this quantitative analysis see Figs. S3 and S4). Fig. 3(e) displays the stack plot of the corresponding MDCs in Fig. 3(d). Our analysis suggests a Fermi velocity of \emph{v}$_{\text{F}} \sim (8.10 \pm 0.75) \times 10^{5}$ ms$^{-1}$ ($5.33 \pm 0.49$ eV$\cdot$\text{\AA}) for these observed metallic linear-like states, larger than that of the topological surface states of Bi$_{2}$Se$_{3}$ \citep{Xia2009}, but smaller than that of the Dirac bands of graphene \citep{Bostwick2007}.

In order to investigate whether these states exhibit two- or three-dimensional dispersion, we perform \emph{k}$_{z}$-dependent measurements of the observed electronic bands by measuring their dispersion upon changing the photon energy $h\nu$ in our ARPES experiments. As can be observed in Fig. 4(a) the metallic states crossing the Fermi level survive at the $\overline{\Gamma}$ point of the BZ at all measured photon energies. Plotting the MDCs at each $h\nu$ in Fig. 4(b) (in the region denoted by the purple rectangular dashed box in the first panel of Fig. 4(a)), we can show that these states lack any dispersion along \emph{k}$_{z}$. Furthermore, based on the circular pocket that these metallic states form on the Fermi surface (see Fig. 2(c)) we estimate a surface carrier density of $\sim$ 3.2 $\times$ 10$^{13}$ cm$^{-2}$ for our samples. Our LEED measurements indicate that no detectable surface reconstruction is present in these crystals within our measurements precision, and thus the possibility of a change in the surface stoichiometry, which may induce new states at the surface due to polarity effects, is ruled out. These evidences, taken collectively, convincingly imply the observation of metallic in-gap states on the surface boundary of Na$_{2}$IrO$_{3}$ at room temperature.

Now we discuss the implications of our experimental observation of Na$_{2}$IrO$_{3}$ electronic bands. We have resolved the electronic band structure of this honeycomb iridate compound at room temperature, and have shown its six-fold symmetry structure along its natural cleavage plane consistent with its previously reported crystal structure in the high temperature non-magnetic phase. Our data indicates that the low-energy physics of this material is governed by metallic in-gap states crossing the Fermi level that have linear-like dispersion, and form a circular Fermi surface pocket at the center of the BZ (the time-reversal invariant Kramers' $\overline{\Gamma}$ point). Photon energy dependence measurements further confirm the 2D nature of these states and provide strong evidence that they are indeed surface states. Our measurements are performed at room temperature, much higher than the temperature at which the antiferromagnetic order sets in. Thus, time-reversal symmetry is preserved in the temperature regime at which our studies are carried out, consistent with the $Z_{2}$ topologically non-trivial phase. In spite of the fact that our experimentally observed bulk bands exhibit a robust gap, our observation of a metallic linear-like surface band feature crossing the Fermi level at the $\overline{\Gamma}$ point provides the first evidence for metallic behavior on the surface boundary of this material, which is in contrast to the theoretical prediction of Dirac cones at the $\overline{M}$ points of the BZ. Therefore, our results suggest that the surface ground state electronic structure of this system is not in a stacked QSH phase as predicted theoretically, but rather reflects an exotic many-body metallic ground state on the surface. The disagreement between our experimental data and theoretical results may reflect the difficultly of modeling strongly correlated systems in theory. These models are typically more accurate for weakly interacting systems rather than strongly correlated ones.

In transport measurements Na$_{2}$IrO$_{3}$ behaves as an insulator. Since transport measurements are mainly governed by bulk states, this further suggests that the observed metallic states in our measurements are surface states. In order to further investigate these observed surface states, more detailed transport measurements are required in the future. The results presented here establish Na$_{2}$IrO$_{3}$ as a rare example of a strongly correlated spin-orbit Mott insulator with surface metallicity. Our study of near-surface electronic structure and the observation of anomalous metallicity reflecting a non-trivial ground state opens up many opportunities for observing exotic transport and Hall effect on the surface of this highly correlated spin-orbit material, and paves the way to future studies exploring the possibility that topological insulators, Weyl semimetals and other topological phases may arise in this class of materials. 

We thank Jonathan D. Denlinger for technical assistance. Work at Princeton University is supported by the US National Science Foundation Grant, NSF-DMR-1006492. M.Z.H. acknowledges visiting scientist support from Lawrence Berkeley National Laboratory and additional partial support from the A. P. Sloan Foundation and NSF-DMR-0819860. G. Cao acknowledges NSF support via grant DMR 1265162. The photoemission measurements using synchrotron X-ray facilities are supported by the Basic Energy Sciences of the US Department of Energy.

\begin{figure}
\centering
\includegraphics[width=14cm]{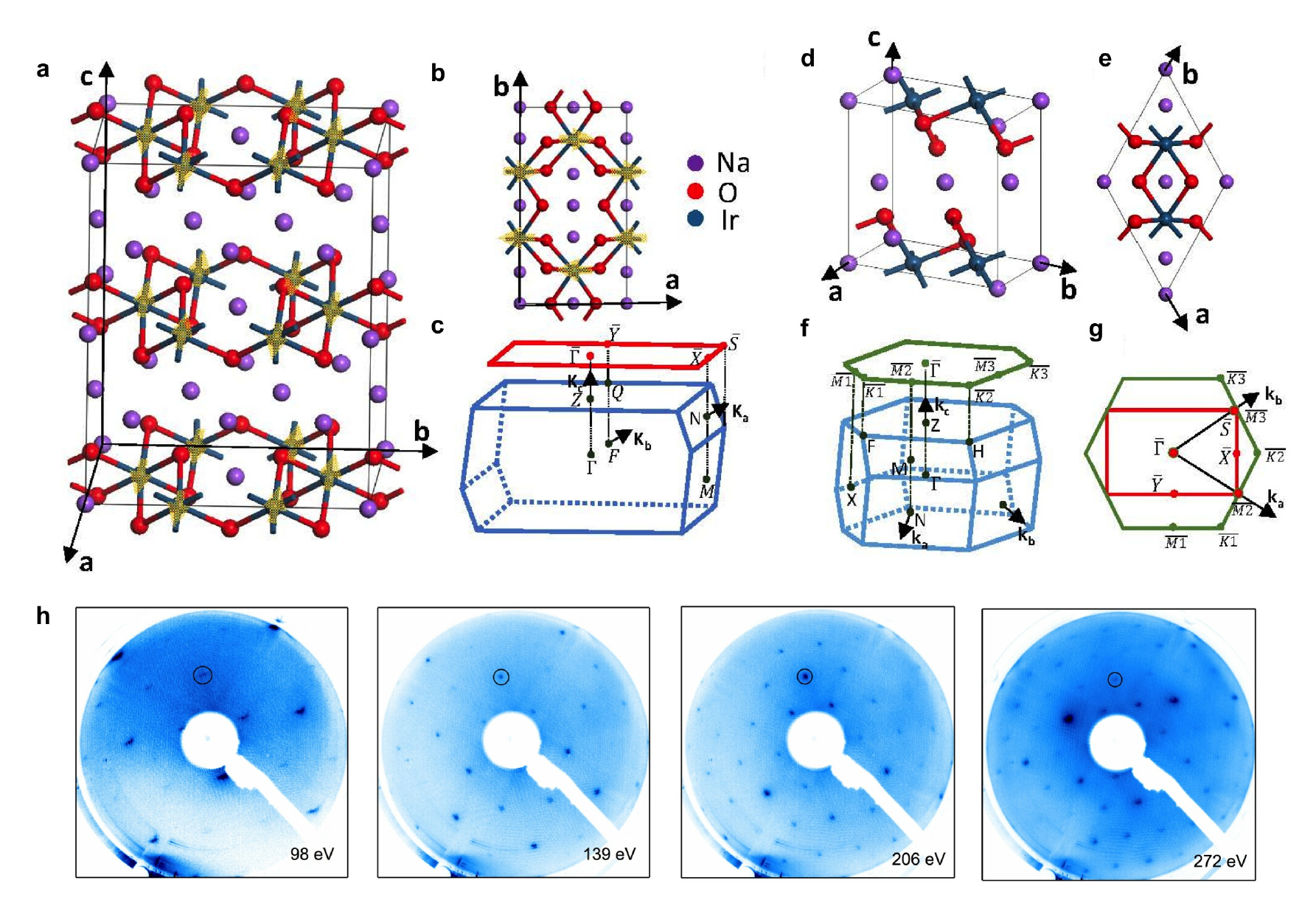}
\caption{(a) Side and (b) top views of the crystal structure and (c) the corresponding Brillouin zone (BZ) of magnetic Na$_{2}$IrO$_{3}$ with the zigzag-AFM spin configuration. The arrows indicate magnetic moments of Ir atoms. (d)-(f) Same as (a)-(c) for non-magnetic Na$_{2}$IrO$_{3}$. The projection of the bulk BZ on the (001) surface is indicated by the red and green rectangles in (c) and (f), respectively. (g) Relationship between the (001) surface BZ of the non-magnetic (green) and zigzag-AFM magnetic (red) Na$_{2}$IrO$_{3}$. (h) Measured low-energy electron diffraction (LEED) patterns of the crystals used in our studies at various electron energies as marked on each panel.}
\end{figure}

\begin{figure}
\centering
\includegraphics[width=14.5cm]{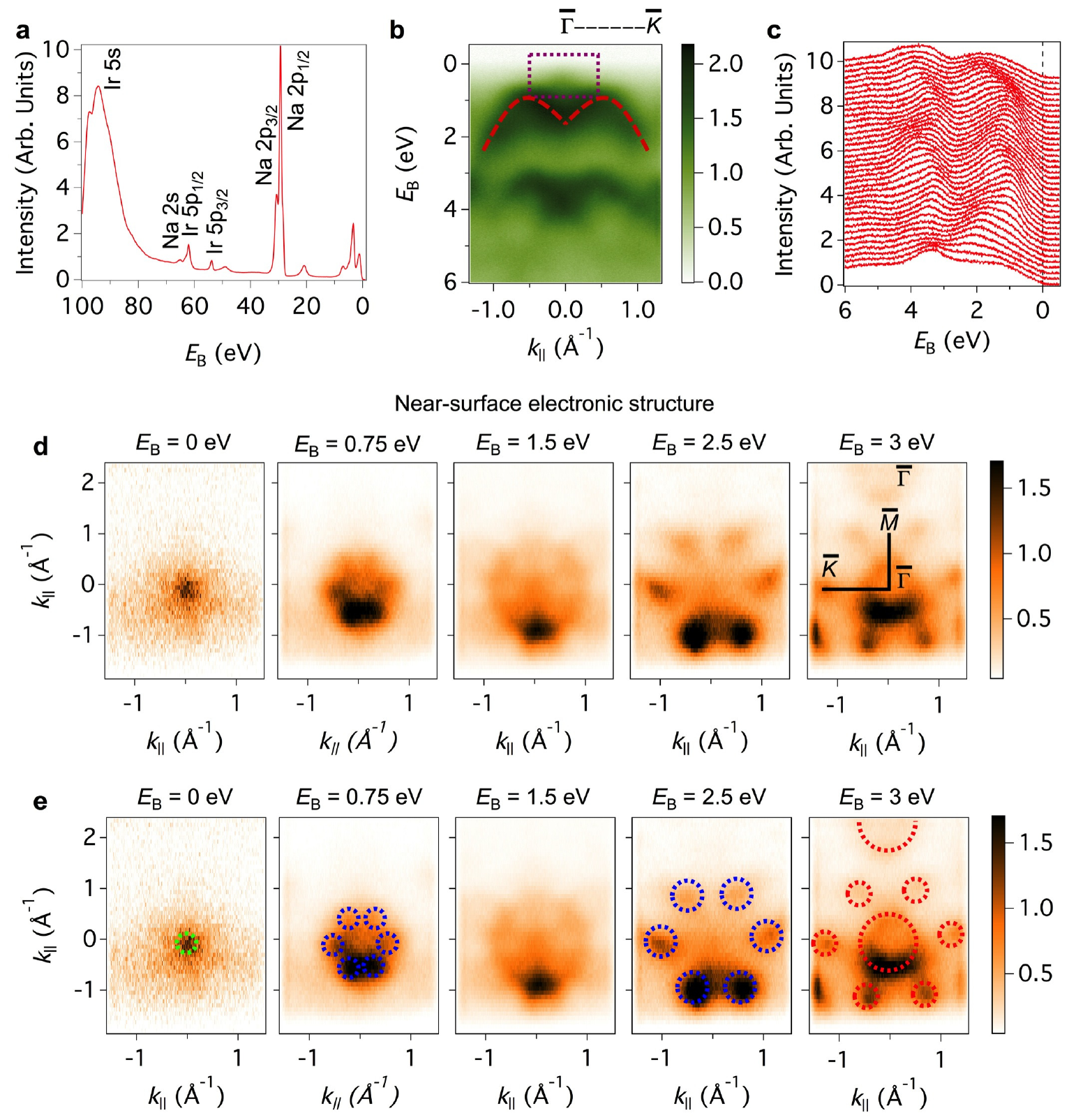}
\caption{(a) Core level spectra of Na$_{2}$IrO$_{3}$. (b) ARPES electronic band structure of Na$_{2}$IrO$_{3}$ along the high symmetry direction $\overline{\Gamma}-\overline{K}$ of the BZ.  (c) Energy distribution curves (EDCs) of the \emph{k}-\emph{E} intensity profile in (b). The dashed line indicates the position of \emph{E}$_{\text{F}}$. (d) Constant binding energy cuts at various binding energies between \emph{E}$_{\text{F}}$ and \emph{E}$_{\text{B}}$ = 3 eV. The corresponding binding energies are indicated at the top of each panel. The high symmetry directions of the BZ, $\overline{\Gamma}-\overline{K}$ and $\overline{\Gamma}-\overline{M}$, are marked on the last panel. (e) Replica of  (d) with dashed guide-to-the-eye lines drawn at the place of each Fermi surface pocket to emphasize the six-fold symmetry of the BZ. Green lines mark the pockets formed by the intensities crossing the \emph{E}$_{\text{F}}$, blue lines the pockets formed by the lower binding energy bulk valence bands, and red lines the pockets formed by the higher binding energy bulk valence bands. These measurements were conducted at \emph{T} = 300 K  with $h\nu$ = 90 eV.}
\end{figure}

\begin{figure}
\centering
\includegraphics[width=14cm]{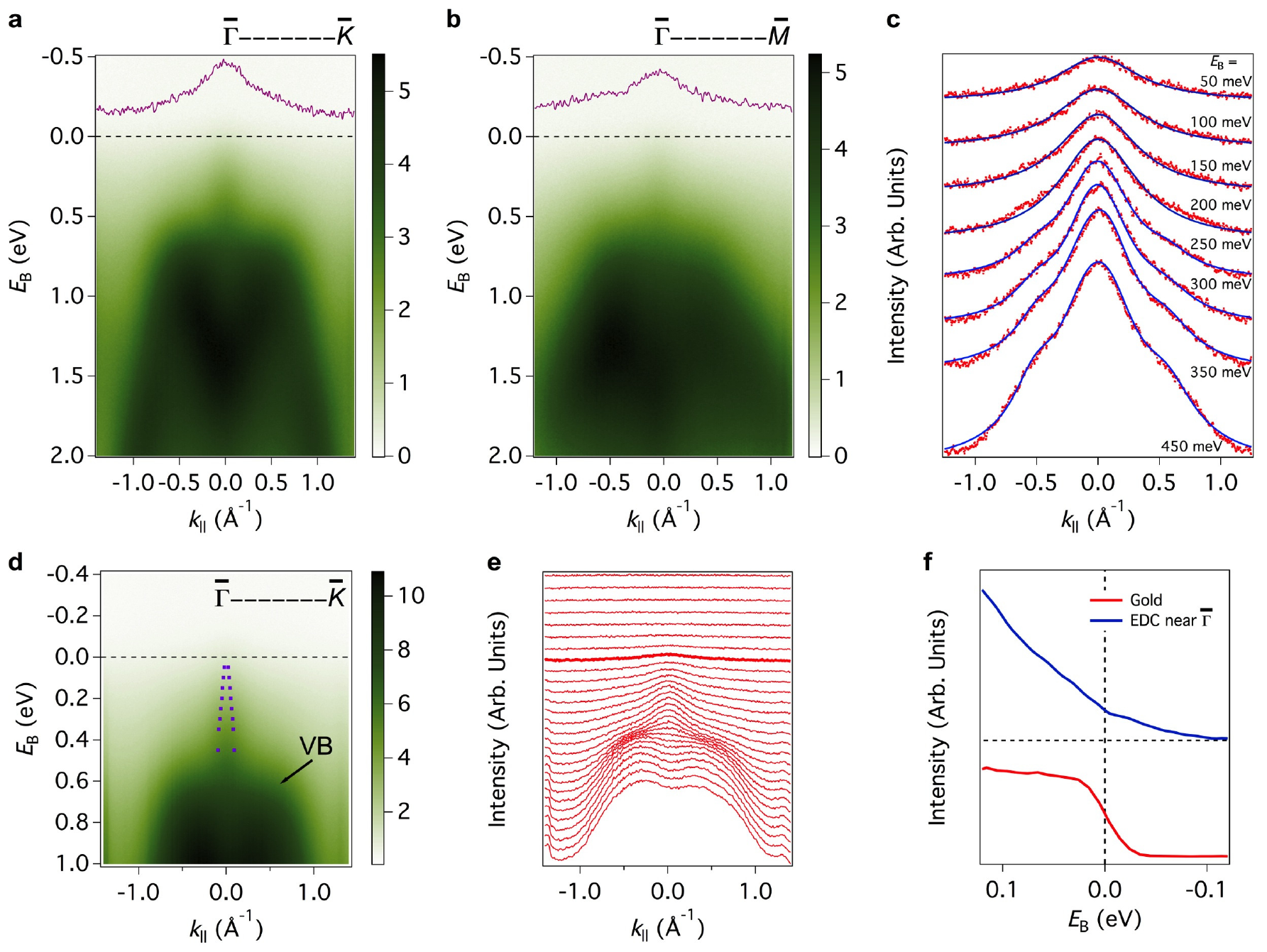}
\caption{ARPES electronic band structure of Na$_{2}$IrO$_{3}$ along the high symmetry directions of (a) $\overline{\Gamma}-\overline{K}$ and (b) $\overline{\Gamma}-\overline{M}$ of the BZ, respectively. The momentum distribution curves (MDCs) at the Fermi level are plotted on top of each panel in purple. (c) MDCs (red) and heuristic fits to the MDCs (blue) of the metallic in-gap states around the $\overline{\Gamma}$ point of the Brillouin zone using two Lorentzians. (d) Zoomed-in version of the metallic in-gap states and (e) their corresponding MDCs. The bold curve in (e) indicates the MDC at \emph{E}$_{\text{F}}$, and the purple dots in (d) show the peak positions of the Lorentzians used to fit to the MDCs of these metallic in-gap state. The horizontal black dashed lines in (a), (b), and (d) denote the position of \emph{E}$_{\text{F}}$. (f) Comparison between the EDCs of polycrystalline gold and that of the Na$_{2}$IrO$_{3}$ samples in our study around the $\overline{\Gamma}$ point of the BZ. Note that the numerical values on the color bars may not be compared across figures, since they are in arbitrary units specific to a certain color plot.} 
\end{figure}

\begin{figure}
\centering
\includegraphics[width=14cm]{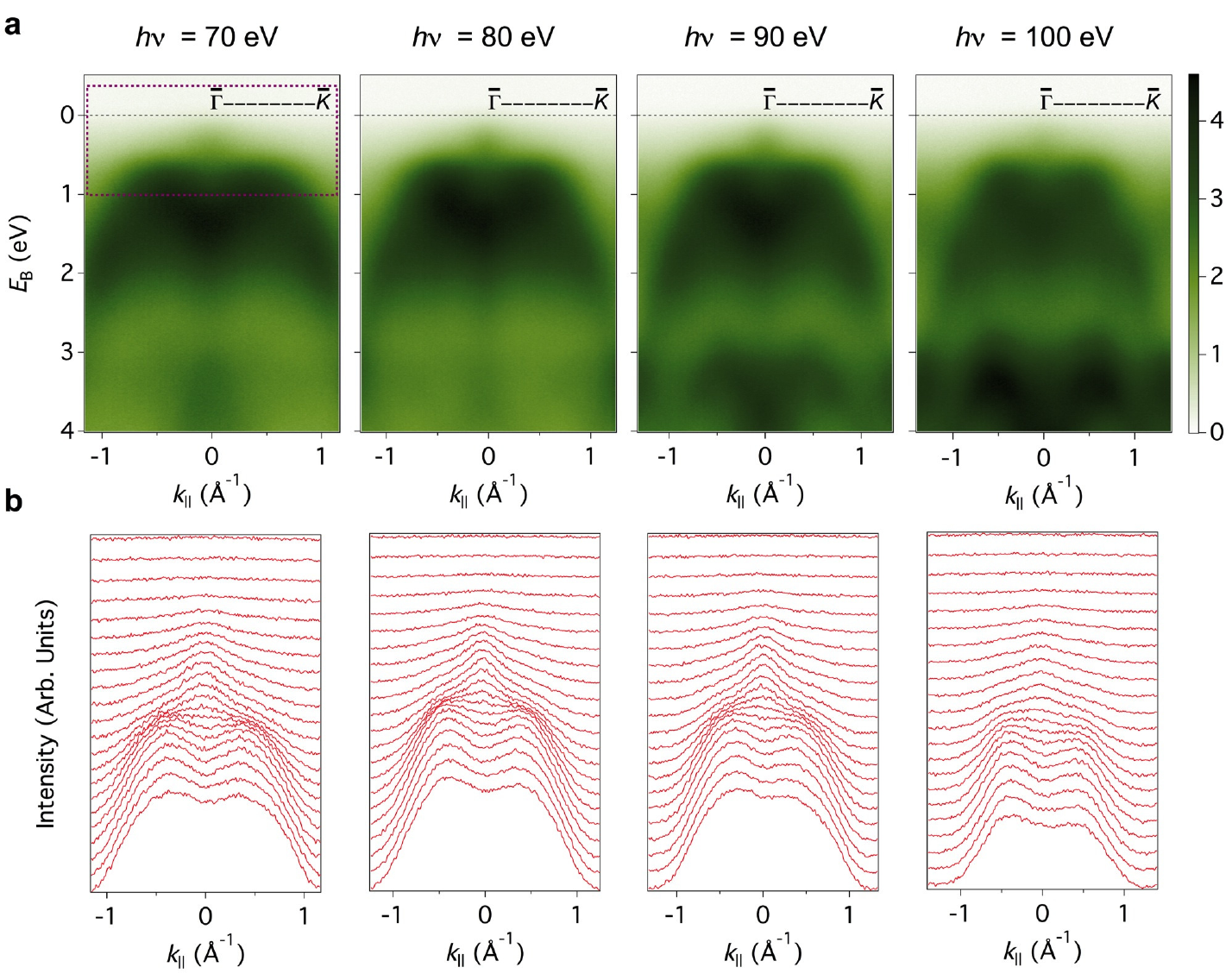}
\caption{(a) Photon energy $h\nu$ ($\propto$ \emph{k}$_{z}$) dependence measurements of the observed metallic linear-like in-gap electronic states and (b) the corresponding MDCs in the region indicated by the purple rectangular dashed box in the first panel of (a). The horizontal black dashed lines in (a) show the position of \emph{E}$_{\text{F}}$.}
\end{figure}

\setcounter{figure}{0} 
\renewcommand{\thefigure}{S\arabic{figure}}

\begin{figure}
\centering
\includegraphics[width=10cm]{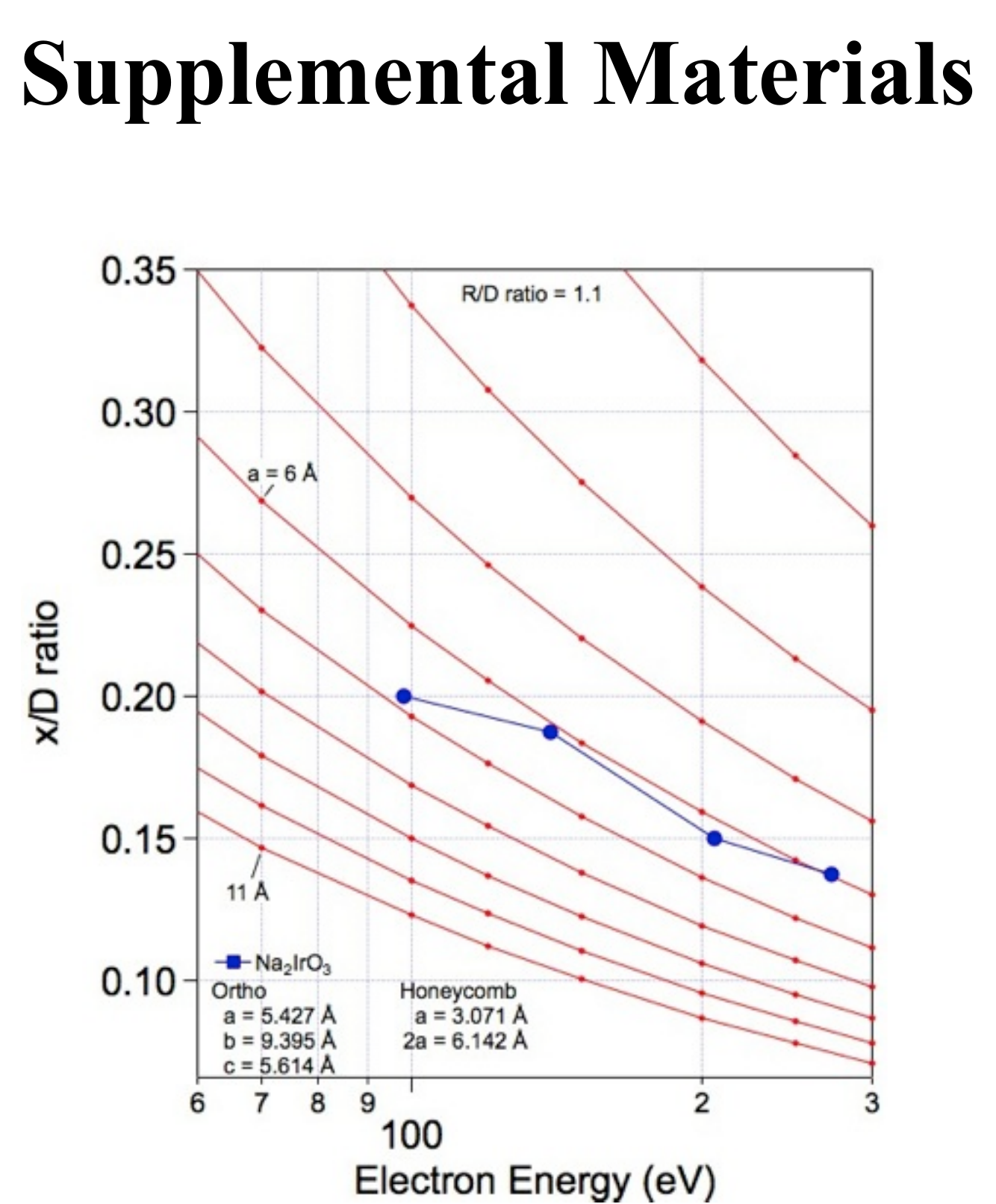}
\caption{Analysis of the low-energy electron diffraction (LEED) data of the same crystals used in our ARPES studies, showing the absence of any detectable surface reconstructions in the Na$_{2}$IrO$_{3}$ samples under study.}
\end{figure}

\begin{figure}
\centering
\includegraphics[width=14cm]{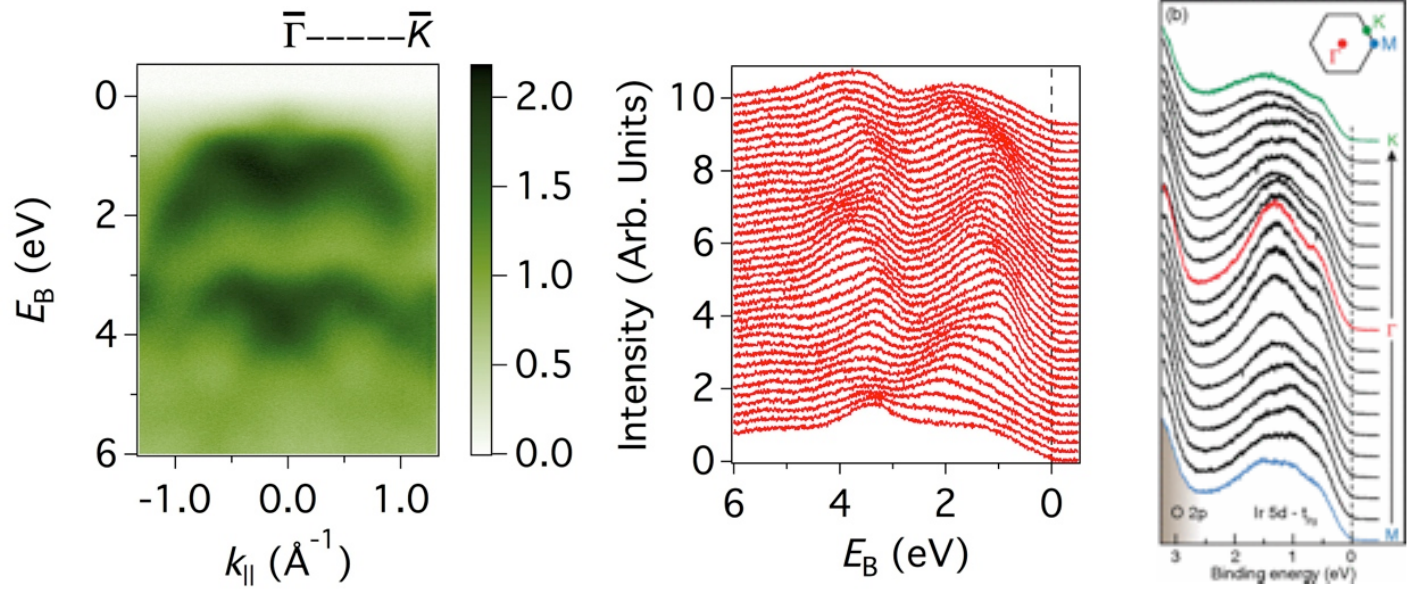}
\caption{Comparison between the bulk bands of Na$_{2}$IrO$_{3}$ resolved in our study (a), and the energy distribution curves (EDCs) reported in \citep{Comin2012} (b), which indicates the qualitative agreement between the observed EDCs of the bulk bands obtained in our study and those reported in \citep{Comin2012}.}
\end{figure}

\begin{figure}
\centering
\includegraphics[width=12cm]{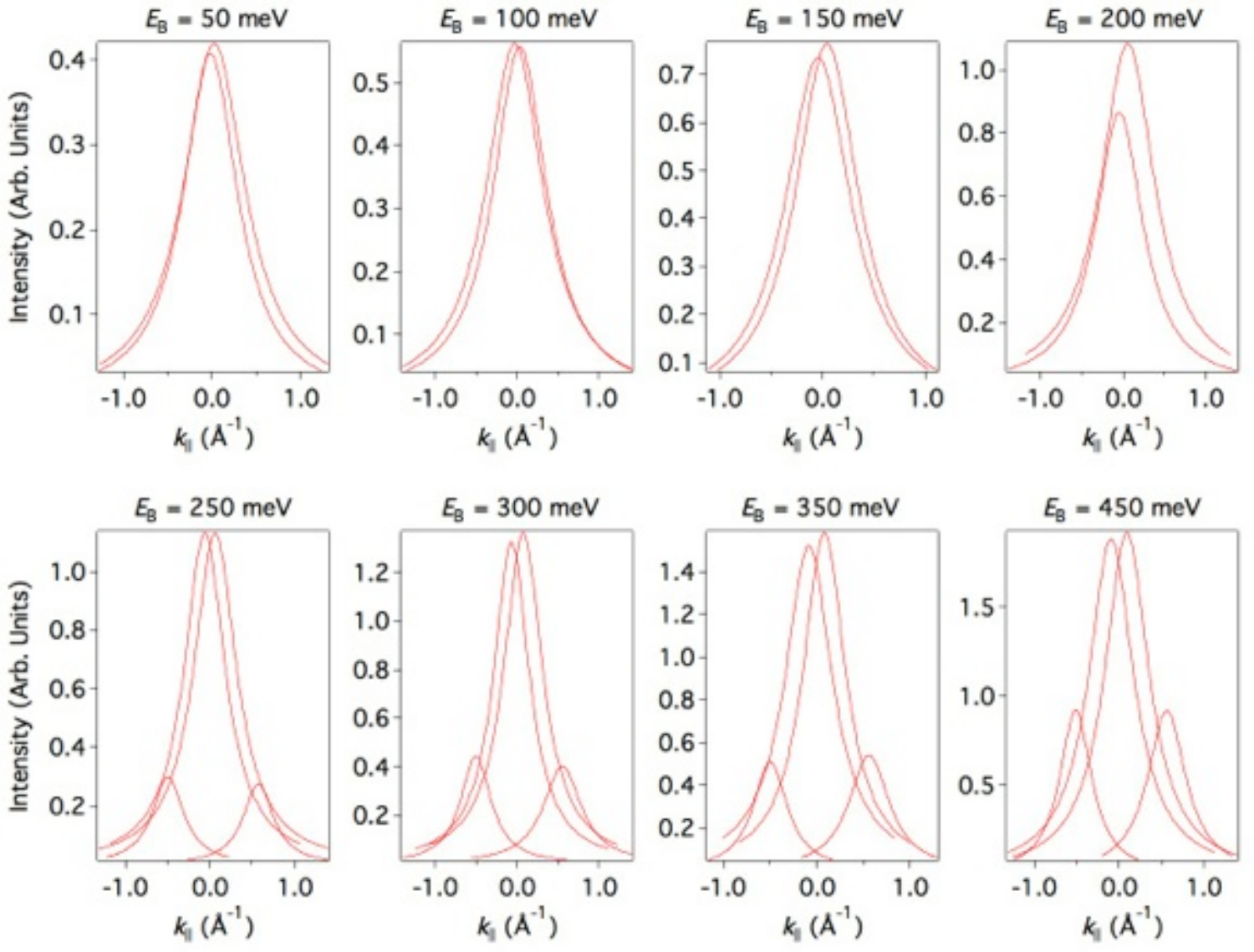}
\caption{Lorentzians used to fit to the MDCs of the metallic in-gap states in the binding energy range between 50 meV and 450 meV around the $\overline{\Gamma}$ point of the Brillouin zone. Note that two additional Lorentzians were used to fit to the tails of the MDCs at higher binding energies to account for some residual intensities from the bulk valence bands.}
\end{figure}

\begin{figure}
\centering
\includegraphics[width=12cm]{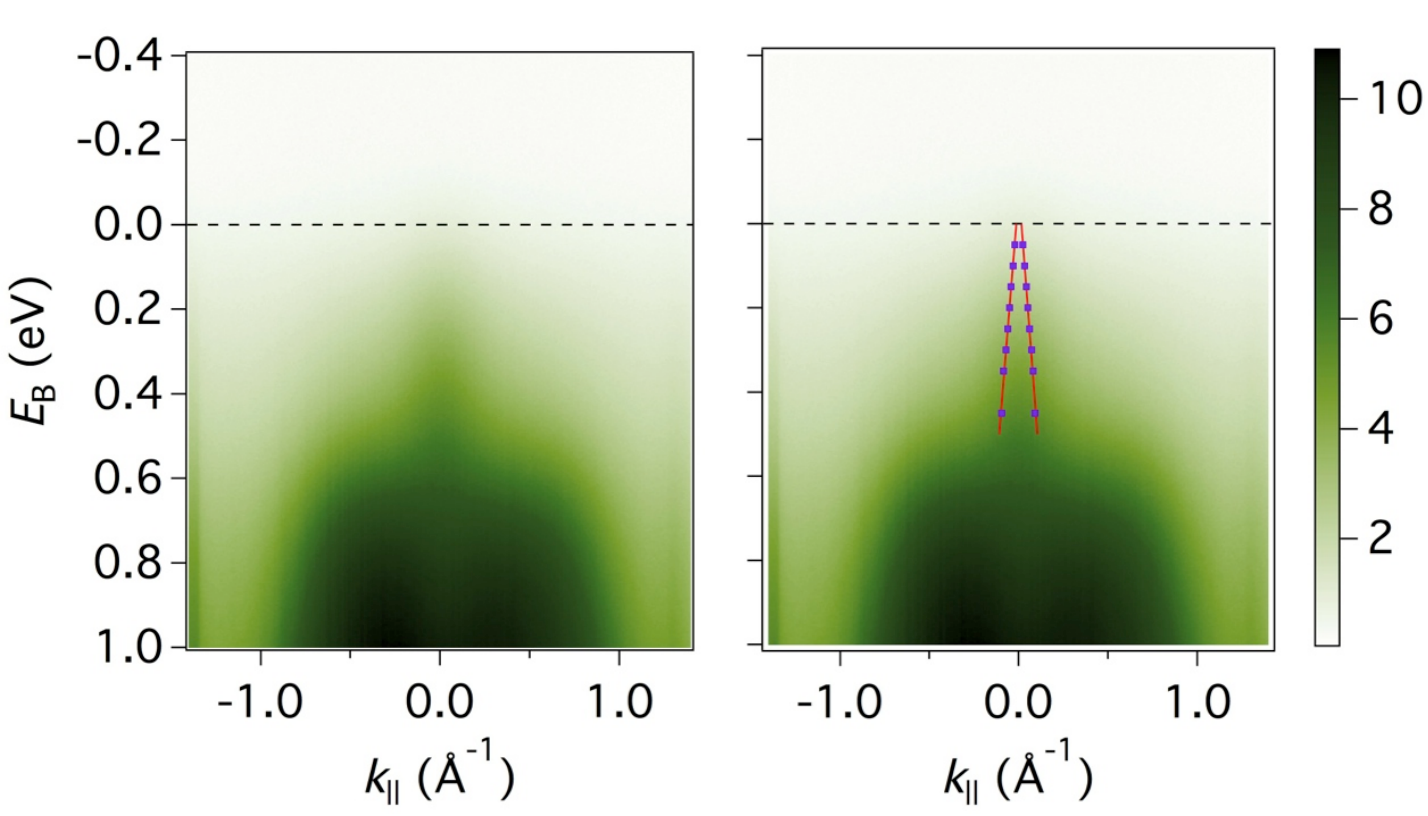}
\caption{Linear fits (red lines) to the peak positions (purple dots) of the Lorentzians used to fit to the MDCs of the metallic in-gap states in the binding energy range between 50 meV and 450 meV around the $\overline{\Gamma}$ point of the Brillouin zone.}
\end{figure}

\end{document}